\documentclass{article}

\usepackage{arxiv}

\usepackage[utf8]{inputenc} 
\usepackage[T1]{fontenc}    
\usepackage{hyperref}       
\usepackage{url}            
\usepackage{booktabs}       
\usepackage{amsfonts}       
\usepackage{nicefrac}       
\usepackage{microtype}      
\usepackage{graphicx}
\usepackage{natbib}
\usepackage{doi}
\usepackage{amsmath}
\usepackage{amssymb,gensymb}
\graphicspath{{./figures/}}

\title{Demonstration of Parametric Instability suppression through optical feedback}

\author{Vladimir Bossilkov\thanks{\texttt{v.bossilkov@gmail.com}}, Jian Liu\thanks{Current affiliations: AEI \& LUH}, Carl Blair, Chunnong Zhao and Li Ju\\
	OzGrav\\
	University of Western Australia\\
	Crawley, Western Australia 6009, Australia\\
}

\renewcommand{\headeright}{}
\renewcommand{\undertitle}{}
\renewcommand{\shorttitle}{}

\hypersetup{
	pdftitle={Demonstration of Parametric Instability suppression through optical feedback},
	pdfsubject={},
	pdfauthor={Vladimir Bossilkov, Jian Liu, Carl Blair, Chunnong Zhao and Li Ju},
	pdfkeywords={},
}

\begin{document}
\maketitle

\begin{abstract}
	We demonstrate the suppression of parametric instability using through optical actuation in an electro-optical feedback loop, stabilising the high order optical mode content in an 80 metre long Fabry-Perot cavity. The loop suppression of the high order mode is achieved by injecting a high order mode with the same frequency and opposite phase. Frequency matching is achieved by measuring the beat note signal between the fundamental and high order mode in transmission of the cavity and applying that signal to an electro-optical modulator to create the required frequency sideband. Spatial mode matching of the sideband to the high order mode is accomplished through the inherent mode overlap between the input injected beam and the high order mode of the cavity. The paper presents the theoretical analysis and experimental demonstration of parametric instability suppression, for an instability which would normally ring up with a parametric gain of approximately 2.5.
\end{abstract}

\section{Introduction}

Advanced gravitational wave detectors, including Advanced LIGO\citep{ligo15} (aLIGO), Advanced Virgo\citep{acernese2015} and KAGRA\citep{aso13}, all use long Fabry-Perot optical cavities which are designed to contain intra-cavity optical power as high as 800 kW. In this paper we explore the suppression of a process called three-mode parametric instability\citep{brag01} (PI), similar to Brillouin Scattering, which scatters a cavity’s fundamental mode into unwanted high order modes, in the process exciting mechanical modes in the test masses of the cavity. Without mitigation many mechanical modes could become unstable due to interactions with many optical modes.

The high order mode beats with the fundamental mode, exerting a radiation pressure force that in turn can further drive the mechanical mode. PI can occur if all of the following conditions are met: the beat frequency between the scattered high order optical mode and the fundamental mode is close to the test mass acoustic mode frequency; there is sufficient spatial overlap between the high order optical mode and the acoustic mode shape; and the system has very low loss. Under these conditions, when operating at a high power, the radiation pressure force exerted on the test mass drives the test mass acoustic mode. If the energy imparted to the mechanical mode surpasses the energy dissipated by the mode it will become unstable.

This phenomenon was first observed in Gingin, Western Australia \citep{zhao2015} and soon after in the aLIGO Livingston Observatory \citep{evans2015}. PI observed during commissioning of aLIGO and had previously limited the operating intra-cavity power\citep{blair2015} and was being mitigated by thermal tuning\citep{degallaix2007} and electrostatic feedback\citep{blair17} control. Recently aLIGO implemented acoustic mode dampers\citep{gras15}, which reduce the quality factor of most unstable mechanical modes by about 10 times, largely mitigating the issue for those detectors.

A number of other PI control methods have been proposed in the past\citep{gras2009,fan2010,miller2011,degallaix2007,blair17}. \citet{zhang10} first proposed injecting a high order optical mode which is matched in frequency and mode shape to a high order mode created by PI, while its phase could be tuned to suppress PI. This concept was demonstrated by \citet{fan2010}, who used optical injection to suppress a high order mode in a Gingin cavity, but that cavity did not have self sustained PI occurring.

We implement a simpler version of the optical injection scheme, and Sec.~\ref{sec:analysis} derives an expression for the change in Parametric Gain due to its implementation. In Sec.~\ref{sec:setupdetails} we present our experimental setup for this control scheme. Sec.~\ref{sec:performance} discusses the specific aspects of the performance of our system, and highlights its effectiveness. Lastly, Sec.~\ref{sec:imp} considers the implementation of this scheme to Gravitational Wave detectors, such as the aLIGO observatories.

\section{Theoretical Derivation}\label{sec:analysis}

We adopt the equation of motion from \citet{dan14}, and consider the feedback of the high order mode as an additional input to the high order mode. Essentially we consider the high order mode to be pumped by feedback amplitude, ${a_1}^{\text{FB}}$:

\begin{equation}\label{eq:differentials}
	\begin{aligned}
		\dot{a_0} =& -\frac{\gamma_0}{2} a_0 + \textbf{i}~G_{\text{01}} b_{\text{m}} a_1 + \sqrt{{\gamma_0}^{\text{in}}}~(A_{p_0} + {a_0}^{\text{in}}),\\
		\dot{a_1} =& -\left(\frac{\gamma_1}{2} + \textbf{i}~\Delta_{\text{m}} \right) a_1 + \textbf{i}~G_{\text{01}} a_0 b_{\text{m}}^{\dagger} \\
		&+ \sqrt{{\gamma_1}^{\text{in}}}~({a_1}^{\text{FB}} + {a_1}^{\text{probe}} + {a_1}^{\text{in}}),\\
		\dot{b_{\text{m}}} =& - \frac{\gamma_{\text{m}}}{2} b_{\text{m}} + \textbf{i}~ G_{\text{01}}~ a_0~ a_1^{\dagger} + \sqrt{\gamma_{\text{m}}}~ b_{\text{th}},
	\end{aligned}
\end{equation}
where: $a_{\text{0,1}}$ is the amplitude of fundamental and high order modes in the cavity, oscillating at frequencies $\omega_0$ and $\omega_1$ respectively; $b_m$ is the amplitude of the mechanical mode oscillating at frequency $\omega_{ m }$; $\gamma_{\text{0,m}}$ is the full linewidth of the fundamental optical and test mass mechanical mode; $G_{\text{01}}$ is the optomechanical coupling strength defined by \citet{dan14}; $A_{p_{0}}$ is the pump amplitude of the fundamental mode; ${a_{\text{0,1}}}^{\text{in}}$ are the vacuum noise fluctuations for the optical modes; and $b_{\text{th}}$ is the thermal noise on the test mass at the mechanical mode frequency; and ${a_1}^{\text{probe}}$ is a probing signal used as an input for transfer function measurements. ${\gamma_1}^{\text{in}}$ is the linewidth of the high order optical mode arising from the tranissivity input test mass. We distinguish this from the linewidth arising from the end test mass tranissivity, ${\gamma_1}^{\text{end}}$, where the sum ${\gamma_1}^{\text{in}} + {\gamma_1}^{\text{out}} + {\gamma_1}^{\text{loss}} = \gamma_1$ is the full linewidth of the high order optical mode, and ${\gamma_1}^{\text{loss}}$ component of the linewidth arising from absorption and scattering losses in the cavity. As $\gamma_0$ is expected to be approximately equal to $\gamma_1$, they can be defined through known test mass parameters:
\begin{equation}\label{eq:gammas}
	\begin{aligned}
		{\gamma_{0,1}}^{\text{in}} =& \frac{c ~ T_1}{2 ~ L},\\
		{\gamma_{0,1}}^{\text{out}} =& \frac{c ~ T_2}{2 ~ L},\\
		{\gamma_{0,1}}^{\text{loss}} =& \frac{c ~\text{P}_\text{loss}}{L},
	\end{aligned}
\end{equation}
where $T_{1,2}$ is the power transmission of the input and end test masses, and $\text{P}_\text{loss}$ is the loss due to laser absorption and scattering per test mass.

$a_{\text{0}}^{\text{in}}$ and $b_{\text{th}}$ are assumed to be zero for this paper, and we consider the small signal regime where we apply the approximation that total circulating cavity power is constant and is equivalent to the fundamental mode power (${a_0}^2 = \bar{n_c} \equiv (2 A_{p_0} / \sqrt{\gamma_0})^2$). Since we treat the fundamental mode as a constant and $b_\text{m}$ and $a_1$ are assumed to be very small, neglect their contribution to the fundamental mode, and thus treat $a_0$ as a constant throughout.

\begin{figure}[h!]
	\centering
	\includegraphics[width=0.45\textwidth]{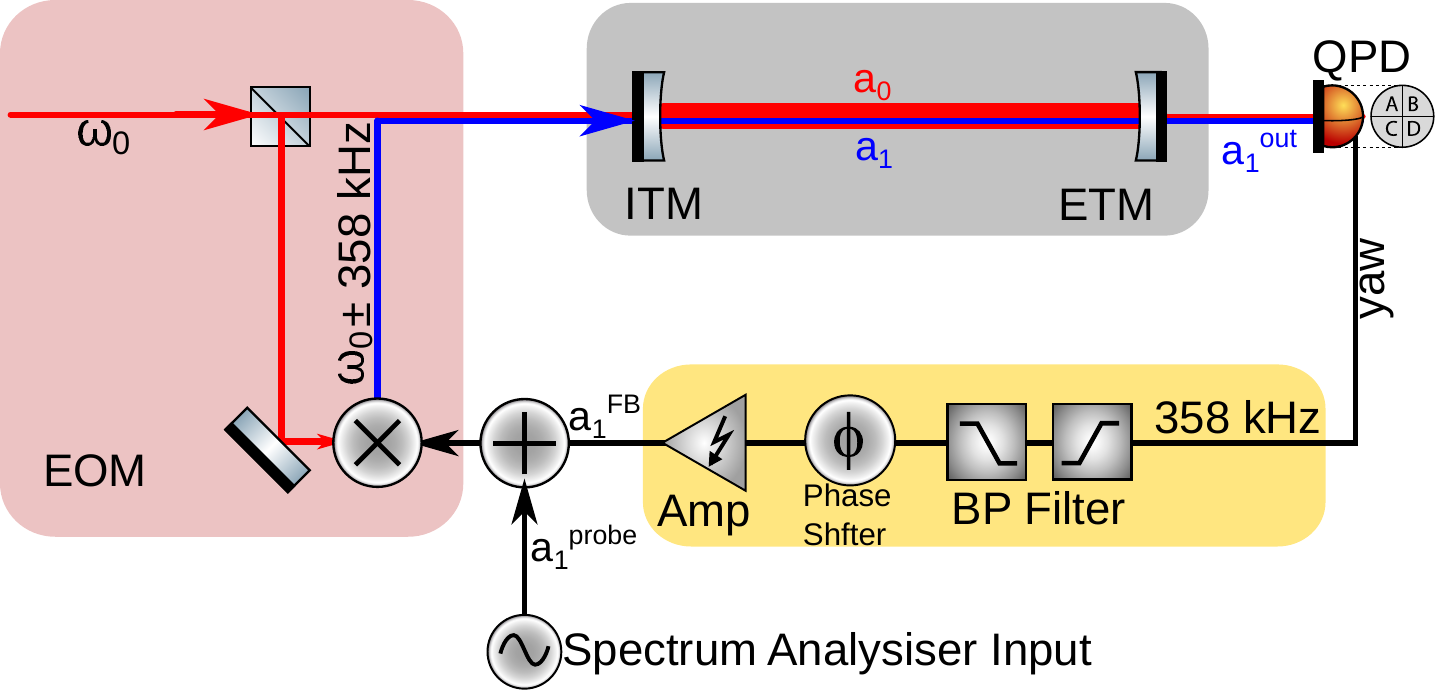}\\
	\caption{Schematic diagram of the function of an EOM injecting a sideband into a cavity, demonstrating the generation of the high order mode for the cavity discussed here. The QPD yaw signal is capable of sensing the beating product between $a_0$ and $a_1$ , and is proportional to just $a_1$ assuming a0 is very large and unchanging. The EOM acts to covert some of the fundamental mode (frequency, $\omega_0$ )to the high order mode (frequency, $\omega_0$ - 358 kHz). In reality the EOM only generates the corresponding sidebands at the high order mode frequency, and the beams’ misalignment with respect to the cavity is what allows the sideband to pump the real high order mode. The band pass filter is sufficiently broad (relative to the optical bandwidths) such that the entire feedback is controlled by a gain defined by the Amplifier, and a phase defined by the Phase Shifter.}
	\label{fig:loop_desc}
\end{figure}

In the absence of a feedback pump for the high order mode (${a_1}^{\text{FB}}$ = 0), these equations can be solved analytically in the frequency domain, taking into account aforementioned assumptions as follows:
\begin{equation}\label{eq:freq_domain}
	\begin{aligned}
		\textbf{i}\Omega~ a_1 (\Omega) =& -\left(\frac{\gamma_1}{2} + \textbf{i}~\Delta_{\text{m}} \right) a_1(\Omega) + \textbf{i}~G_{\text{01}}  a_0 b_{\text{m}}^{\dagger}(\Omega)\\
		&+ \sqrt{{\gamma_1}^{\text{in}}} ({a_1}^{\text{probe}} + {a_1}^{\text{in}}),\\
		\textbf{i}\Omega~ b_{\text{m}} (\Omega) =& - \frac{\gamma_{\text{m}}}{2} b_{\text{m}}(\Omega)+ \textbf{i}~ G_{\text{01}}~ a_0~ a_1^{\dagger}(\Omega),
	\end{aligned}
\end{equation}
with the solution to the high order optical mode within the cavity being:
\begin{equation}\label{eq:openloopTFs}
	\begin{aligned}
				\frac{a_1 (\Omega)}{{a_1}^{\text{probe}}(\Omega) + {a_1}^{\text{in}}(\Omega)} =& \frac{\sqrt{{\gamma_1}^{\text{in}}} (\frac{\gamma_{m}}{2}+ \textbf{i} ~\Omega)}{(\frac{\gamma_{m}}{2}+ \textbf{i} ~\Omega)(\frac{\gamma_{1}}{2} + \textbf{i}~\Delta_{\text{m}}+ \textbf{i}~\Omega)- G_{\text{01}}^2~\bar{n_c}},\\
	\end{aligned}
\end{equation}
where $\Omega$ is the observation Fourier frequency, with $\Omega = 0$ corresponding to the mechanical mode frequency (in our case 357.9 kHz). The quantity ${a_1}^{\text{out}} /{a_1}^{\text{probe}}$ is a transfer function of the high order mode cavity gain measured at the end test mass transmission port, and can be measured with a swept sine probe as ${a_1}^{\text{probe}} \gg {a_1}^{\text{in}}$ to measure the response across this optical mode for an open loop measurement, as per our Fig.~\ref{fig:loop_desc}, with ${a_1}^{\text{FB}} =0$. Cavity gain for the high order mode, $G_{\text{cav}}$, is:
\begin{equation}\label{eq:cavgain}
	\begin{aligned}
		G_{\text{cav}} =& \frac{{a_1}^{\text{out}} (\Omega)}{{a_1}^{\text{probe}}(\Omega)} = \frac{\sqrt{{\gamma_1}^{\text{out}}} a_1 (\Omega)}{{a_1}^{\text{probe}}(\Omega)} \\
		=&\frac{\sqrt{{\gamma_1}^{\text{in}} {\gamma_1}^{\text{out}}} (\frac{\gamma_{m}}{2}+ \textbf{i} ~\Omega)}{(\frac{\gamma_{m}}{2}+ \textbf{i} ~\Omega)(\frac{\gamma_{1}}{2} + \textbf{i}~\Delta_{\text{m}}+ \textbf{i}~\Omega)- G_{\text{01}}^2~\bar{n_c}}.
	\end{aligned}
\end{equation}

We can obtain the stability criterion of the system by solving for the poles of transfer function in Eq.~(\ref{eq:openloopTFs}), which is consistent with that derived by \citet{dan14}:
\begin{equation}\label{eq:pigain_simple}
	\text{R}_{0} = \frac{4 G_{\text{01}}^2~\bar{n_c}}{\gamma_{1} \gamma_{ m }} \geq 1 + \left(\frac{2 \Delta_{\text{m}}}{\gamma_{ m } + \gamma_1 }\right)^2,
\end{equation}
where, $\text{R}_{0}$ is the parametric gain of the interaction. When $\text{R}_{0}$ is larger than the right hand side of this inequality, parametric instability would occur.

Fig.~\ref{fig:loop_desc} schematically illustrates the generation of the sideband which produces the high order mode necessary to prevent PI. The feedback loop alters the phase such that a new high order mode at the same frequency is injected (also at the lower sideband frequency) and can destructively interfere with the existing high order mode. In relation to the PI feedback loop presented by \citet{evans2015}, this process is capable of eliminating the build-up of undesirable high order mode content and prevent further excitation of a mechanical mode by reducing the radiation pressure applied to the test mass.

In order to manipulate this interaction, we define the feedback, discussed in Sec.~\ref{sec:setupdetails}, as the complex quantity $G_\text{e}$. With the feedback loop implemented, the high order mode is effectively pumped with a term ${a_1}^{\text{FB}} = G_\text{e} {a_1}^{\text{out}} = G_\text{e} \sqrt{{\gamma_1}^{\text{out}}} a_1$, where $G_\text{e}$ describes the gain from electronics in the feedback loop. Now in frequency domain, we rewrite the equation for $a_1$  from Eq.~(\ref{eq:differentials}) to see how it changes in comparison to the solution in Eq.~(\ref{eq:freq_domain}):
\begin{equation}\label{eq:freq_a1}
	\begin{aligned}
		\textbf{i}\Omega~ a_1 (\Omega) =& -\left(\frac{\gamma_1}{2} + \textbf{i}~\Delta_{\text{m}} \right) a_1(\Omega) + \textbf{i}~G_{\text{01}}  a_0(\Omega) b_{\text{m}}^{\dagger}(\Omega)\\
		&+ \sqrt{{\gamma_1}^{\text{in}} {\gamma_1}^{\text{out}}}~G_\text{e}~a_1(\Omega) + \sqrt{{\gamma_1}^{\text{in}}} ({a_1}^{\text{probe}}(\Omega) + {a_1}^{\text{in}}(\Omega))\\
		=&-\left(\frac{\gamma_1'}{2} + \textbf{i}~\Delta_{\text{m}}' \right) a_1(\Omega) + \textbf{i}~G_{\text{01}}  a_0(\Omega) b_{\text{m}}^{\dagger}(\Omega)\\
		&+ \sqrt{{\gamma_1}^{\text{in}}} ({a_1}^{\text{probe}}(\Omega) + {a_1}^{\text{in}}(\Omega)),\\
	\end{aligned}
\end{equation}
where, $\gamma_1' = \gamma_1 - 2 \sqrt{{\gamma_1}^{\text{in}} {\gamma_1}^{\text{out}}}~\text{Re}[G_\text{e}]$ and $\Delta_{\text{m}}' = \Delta_{\text{m}}+\sqrt{{\gamma_1}^{\text{in}} {\gamma_1}^{\text{out}}}\text{Im}[G_\text{e}]$ are effective changes in high order mode optical linewidth and PI detuning for a closed loop system. Using Eq.~(\ref{eq:freq_a1}) gives the solution for the cavity gain:
\begin{equation}\label{eq:openloopmod}
	\begin{aligned}
		G_{\text{cav}} = \frac{\sqrt{{\gamma_1}^{\text{out}}} a_1 (\Omega)}{{a_1}^{\text{probe}}(\Omega) + {a_1}^{\text{in}}(\Omega)} =& \frac{\sqrt{{\gamma_1}^{\text{in}} {\gamma_1}^{\text{out}}} (\frac{\gamma_{m}}{2}+ \textbf{i} ~\Omega)}{(\frac{\gamma_{m}}{2}+ \textbf{i} ~\Omega)(\frac{\gamma_{1}'}{2} + \textbf{i}~\Delta_{\text{m}}'+ \textbf{i}~\Omega)- G_{\text{01}}^2~\bar{n_c}}.\\
	\end{aligned}
\end{equation}

The same solution can be derived through control theory. In an open loop:
\begin{equation}
	G_{\text{open}} = G_\text{e} G_{\text{cav}},
\end{equation}
and in a closed loop where ${a_1}^{\text{FB}} = G_\text{e} {a_1}^{\text{out}}$:
\begin{equation}
	\frac{{a_1}^{\text{FB}}}{{a_1}^{\text{probe}}} = \frac{G_\text{open}}{1-G_\text{open}},
\end{equation}
yields the same solution as Eq.~(\ref{eq:openloopmod}).

${a_1}^{\text{FB}} / {a_1}^{\text{probe}}$ is measured in a experimentally in Sec.\ref{sec:performance}, assuming a large enough ${a_1}^{\text{probe}}$ to render
${a_1}^{\text{in}}$ negligible.

The instability criterion may once again be evaluated:
\begin{equation}\label{eq:pigain_feedback}
	\begin{aligned}
		\text{R}_{0}' = \frac{4 G_{\text{01}}^2~\bar{n_c}}{\gamma_{1}' \gamma_{ m }} = \text{R}_{0} \left(\frac{\gamma_{1}}{\gamma_{1}'} \right) \implies \text{R}_{0} \geq \left(\frac{\gamma_{1}'}{\gamma_{1}} \right) \left(1 + \left(\frac{2 \Delta_{\text{m}}'}{\gamma_{ m } + \gamma_{1}'}\right)^2 \right).	
	\end{aligned}
\end{equation}

Based on the change in the instability criterion arising from the feedback term (comparing equations Eq.~(\ref{eq:pigain_simple}) and Eq.~(\ref{eq:pigain_feedback})), we derive a term for effective PI gain reduction factor, $\eta$, for cases where $\Delta_{\text{m}} = 0$, where parametric gain would be maximised for a specific mechanical mode. Furthermore, since the $\gamma_1 \gg \gamma_{\text{m}}$, the expression can be simplified to:
\begin{equation}\label{eq:pi_reduction}
	\eta=\left(1-\text{Re}[\xi]\right) \left(1+\left(\frac{ \text{Im}[\xi]}{1 - \text{Re}[\xi]}\right)^2 \right) = \frac{(1-\xi)(1-\xi^{\dagger})}{1-\text{Re}[\xi]},
\end{equation}
where $\xi = 2\sqrt{{\gamma_1}^{\text{in}} {\gamma_1}^{\text{out}}} G_\text{e} / \gamma_1$.

As a consequence of the leading $(1-\text{Re}[\xi])$ term in Eq.~(\ref{eq:pi_reduction}), feedback would suppress PI where $G_\text{e}$ is negative, or equivalently having a phase of [$\frac{\pi}{2}$,$\frac{3\pi}{2}$]. A larger suppression factor is possible near those boundary conditions due to the feedback phase significantly contributing to effective detuning (through the $\text{Im}[\xi]$ term). Electrostatic damping and acoustic mode damper strategies for PI mitigation, function only to increase the mechanical loss factor ($\gamma_{ m }$), and in comparison, an obvious advantage to using optical feedback is that far greater suppression can be achieved since the effective detuning, $\Delta_{\text{m}}'$, can be controlled through tuning the feedback phase.

\section{Experimental Setup Overview}\label{sec:setupdetails}

The Gingin facility has been used for PI experiments since 2008\citep{zhao2008}. Table~\ref{tab:gingincavityparameters} summarises the current cavity parameters that are relevant for this experiment.

In contrast to the previous implementation\citep{fan2010} of optical feedback, where a separate out-of-phase high order mode was produced, we only generated the frequency matched sideband through phase modulation of the carrier mode. Due to an inherent input misalignment to the cavity, some of this sideband couples to a cavity high order mode. The method presented could provide an alternative way to suppress parametric for gravitational wave observatories, and should also be applicable to controlling other optomechanical systems where high order modes are problematic.

\begin{table}[h]
	\caption{Relevant Gingin cavity parameters}
	\centering
	\begin{tabular}{c | c } 
		\hline
		Parameters & Value \\
		\hline
		Mass of test mass, $m$& 0.8 kg \\
		Test mass Q-factor & $2 \times 10^{6}$ \\
		Cavity length, $L$  & 74 m \\
		ITM, ETM radii of curvature, $R_{1}$, $R_{2}$ & 40.0 m, 40.0 m \\
		ITM, ETM mirror power transmissivity, $T_{1}$, $T_{2}$ & 200 ppm,  25 ppm\\
		Test Mass absorption/scattering loss, $\text{P}_\text{loss}$& 110 ppm \\
		Laser wavelength, $\lambda$ & 1064 nm \\
		Laser injection power, $\text{P} _ { \rm { in } }$ & 5 W \\
		Intra-cavity power, $\bar{n_c}$ & 13 kW \\
		Mechanical Mode frequency, $\omega_{ m }$ & 2$\pi$ 357.9 krad/s\\
		Total Optical linewidth, $\gamma_{0,1}$ & 2$\pi$ 143.47 rad/s\\
		Input Optical linewidth, ${\gamma_{0,1}}^{\text{in}}$ & 2$\pi$ 8.06 rad/s\\
		Transmission Optical linewidth, ${\gamma_{0,1}}^{\text{out}}$ & 2$\pi$ 64.48 rad/s\\
		Loss Optical linewidth, ${\gamma_{0,1}}^{\text{loss}}$ & 2$\pi$ 70.93 rad/s\\
		\hline
	\end{tabular}
	\label{tab:gingincavityparameters}
\end{table}

\begin{figure*}
	\centering
	\includegraphics[width=0.85\textwidth]{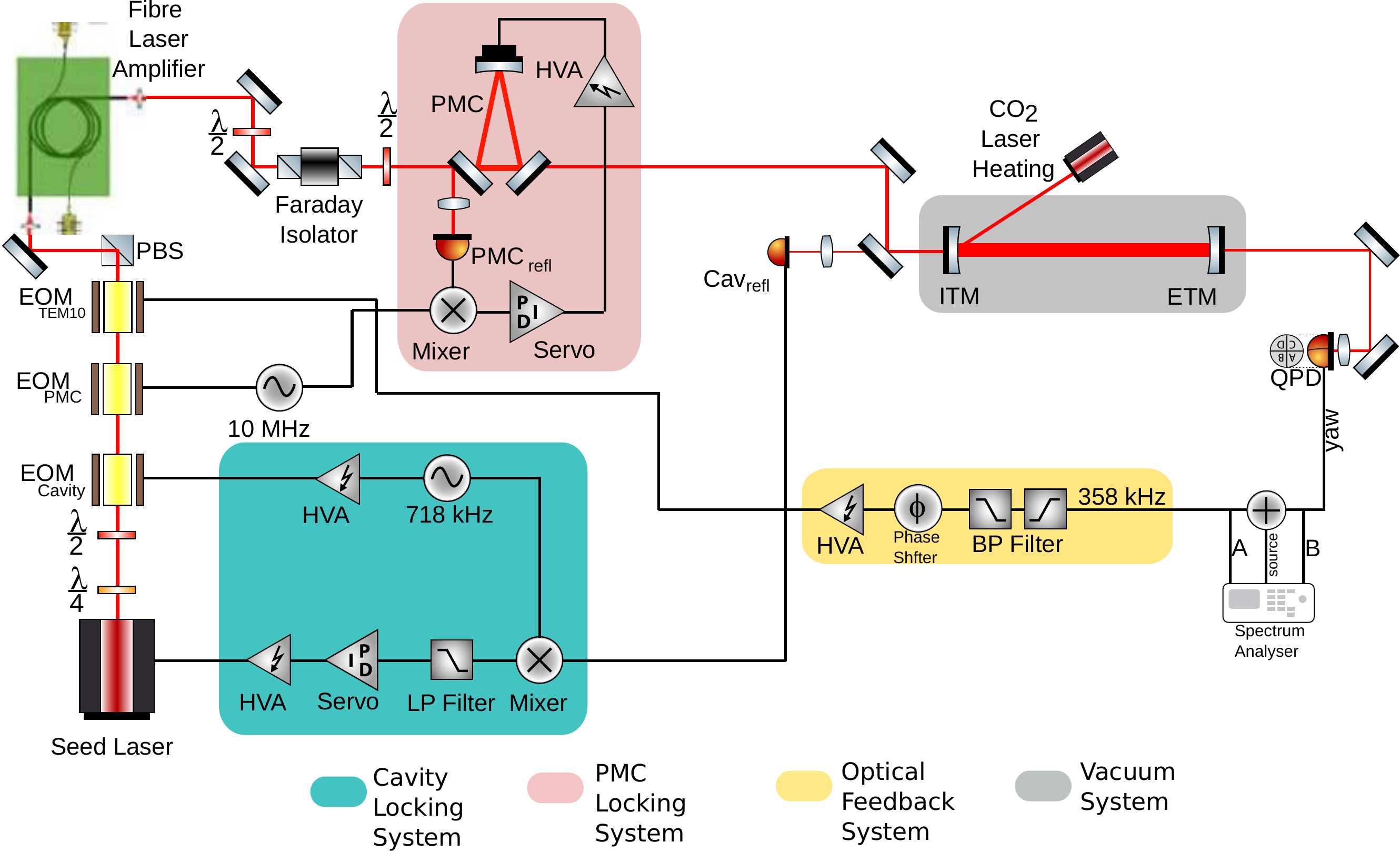}\\
	\caption{Setup of the Gingin East Cavity, and its sub-systems. The cavity locking loop employs PDH locking \protect\citep{drever83},in blue, to lock the laser to the cavity. The Fibre Laser Amplifier is capable of outputting laser power between 1~W and 50~W. The optical feedback loop, in yellow, is responsible for the injection of sideband which we use to suppress PI. The pre-mode cleaner (PMC) loop, in pink, ensures a clean Gaussian input for our cavity, and our injected sidebands are within the bandwidth of this mode cleaner. Most of our feedback loops use PID controllers as the servo, and are further amplified by High Voltage Amplifiers (HVA) to provide sufficient dynamic range. The CO$_2$ laser is used to assist in controlling the ITM radius of curvature to adjust cavity mode spacing. The functionality of the Optical Feedback system is discussed throughout. The spectrum analyser in this configuration can be used to measure the open loop transfer function in a close loop, by measuring $B/A$ with a swept sine source signal.}
	\label{fig:expsetup}
\end{figure*}

The overall set-up is illustrated in Fig.~\ref{fig:expsetup}. The key readout component, the Quadrant Photodiode (QPD), is placed at the transmission port of the cavity. The `yaw' signal senses the beat note power arising from beating between the fundamental, $\text{TEM}_{00}$, and first order optical mode, $\text{TEM}_{10}$, of the cavity. The QPD yaw signal is sensitive to both the $\text{TEM}_{10}$ and $\text{TEM}_{01}$ modes because the cavity axes are not aligned with the QPD splitting axes, and the modes themselves are of different frequencies due to test mass imperfections breaking degeneracy between the two first order modes. You can see the two beat notes as peaks in Fig.~\ref{fig:spect} and \ref{fig:tf}. We know the lower frequency beat note is the $\text{TEM}_{10}$ as we have imaged it with a phase-camera-like set-up, and the other is inferred to be $\text{TEM}_{01}$.

The signal on the QPD is filtered to suppress unwanted noise, amplified and applied to an Electro-Optic Modulator (EOM), $\text{EOM}_{\text{TEM10}}$ , with the phase shifter acting to optimise the stability of the feedback by controlling the phase. The lower sideband generated by the EOM will be a frequency matched to the high order mode participating in PI.

The power loss attributable to the test masses, $\text{P}_\text{loss}$ is calculated as the required loss per test mass to yield the correct measured cavity Finesse\cite{fang21}. This loss figure (stated in Table~\ref{tab:gingincavityparameters}) is used in Eq.~(\ref{eq:gammas}), to give the $\gamma_1$ term in the table. There should be more scatter loss for the high order optical modes but this difference is small for the first order optical mode (compared to the fundamental mode), and is consequently neglected.

The end mirror of our cavity has a mechanical mode at 357.9kHz, with a Q-factor of approximately $2 \times 10^{6}$, that has a known strong opto-mechanical interaction. The $\text{TEM}_{10}$ mode can be thermally tuned such that the frequency difference between it and the fundamental mode is 357.9~kHz, using a $\text{CO}_2$ laser to alter the radius of curvature of the ITM. The mechanical Q-factor was measured by observing the time constant of the ring-down of the mechanical motion of the test mass after suddenly unlocking the cavity which had been sustaining a saturated PI. This signal could be observed in the lever arm of the angular control system. The angular control system provides local angular control with a bandwidth of about 6~Hz, which has little effect on the signal at 357.9~kHz.

\section{Performance}\label{sec:performance}

In the absence of the feedback loop, the dominant high order mode pumping mechanism in our cavity is the input misalignment coupling\cite{anderson84} due to added phase delays in the cavity round-trip photon time. A small fraction of the input power is coupled to the high order mode as not all input power is perfectly aligned to resonate with the fundamental mode. Our feedback loop takes advantage of directly pumping the high order mode side-band, essentially taking advantage of the same process \citet{anderson84} uses to control alignment. The difference is that we use the signal to directly alter the high order mode content, rather than change the alignment. This injection method has much higher coupling to the high order mode than the coupling from the fundamental mode and we provide an estimate for this coupling fraction in this section, and thus transfer function measurements assume that the injection signal contribution to ${a_1}^{\text{probe}}$ greatly dominates any other effect occurring within the cavity (${a_1}^{\text{probe}} \gg {a_1}^{\text{in}}$). If this assumption does not hold, coherent transfer function measurements would not be possible.

\begin{figure}[h!]
	\centering
	\includegraphics[width=0.65\textwidth]{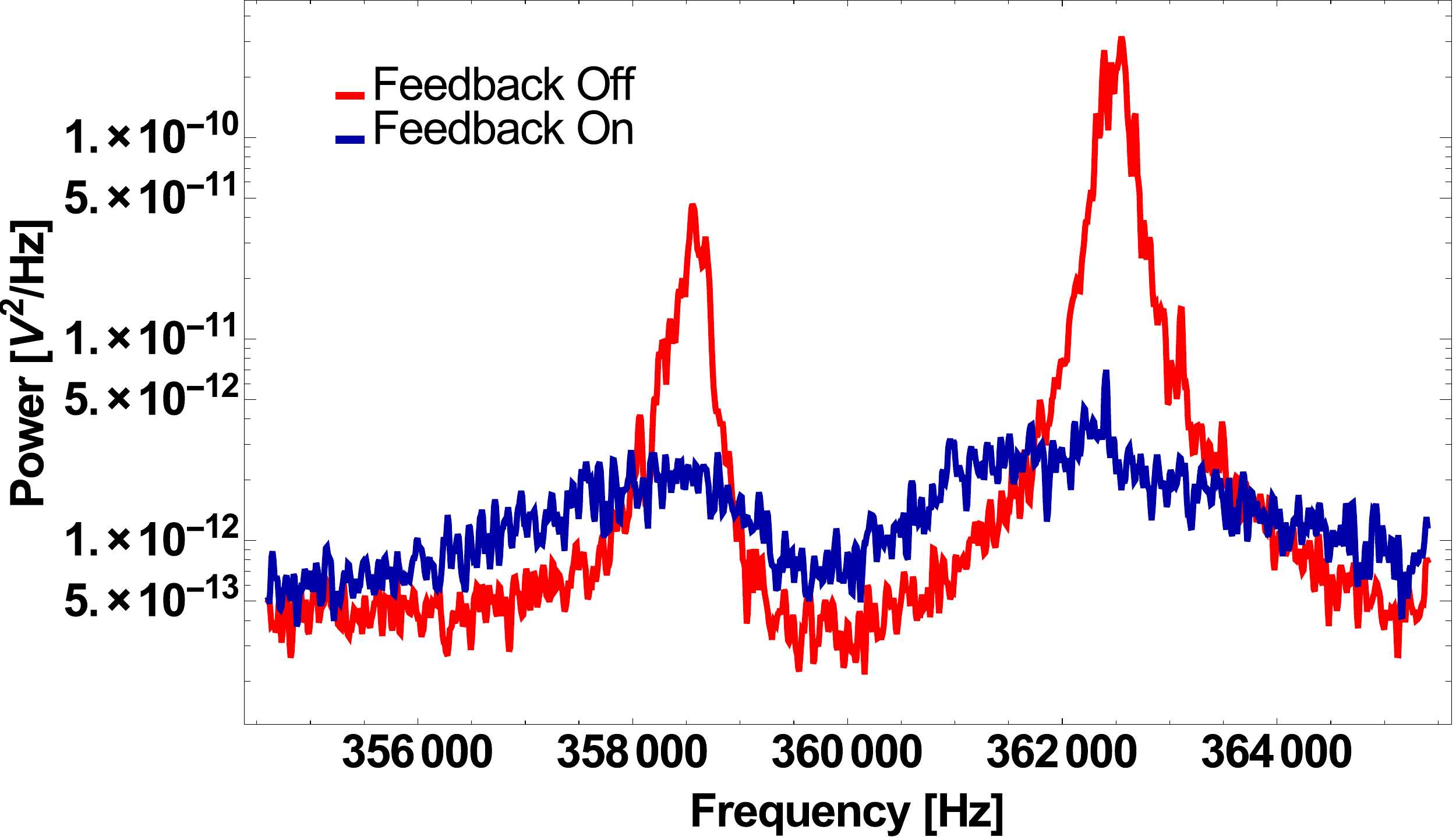}\\
	\caption{Power Spectral Density of the beat notes generated from beating with the fundamental mode and the two first order modes, which are observed at the QPD at the cavity transmission port. The amplitudes of these peaks are proportional to the radiation pressure exerted on test masses in the PI process due to the corresponding optical modes. The feedback loop acts to suppress this beating in the cavity. The first peak corresponds to beating with fundamental mode ($\text{TEM}_{00}$) and the the $\text{TEM}_{10}$ mode, and the other is beating with the $\text{TEM}_{01}$ mode. We show that the amplitude of both optical modes is reduced by the feedback loop.}
	\label{fig:spect}
\end{figure}

Fig.~\ref{fig:spect} illustrates the difference in output observed by the QPD (illustrated in Fig.~\ref{fig:expsetup}), with the feedback loop on (red) and off (blue), highlighting a change in effective optical linewidth that arises from the feedback. The two peaks represent the beat note between $\text{TEM}_{00}$ and $\text{TEM}_{10}$ modes (left peak), and the beat note between $\text{TEM}_{00}$ and $\text{TEM}_{01}$ mode (right peak). The apparent suppression of the high order modes is clear when the feedback loop is turned on.

\begin{figure}[h!]
	\centering
	\includegraphics[width=0.65\textwidth]{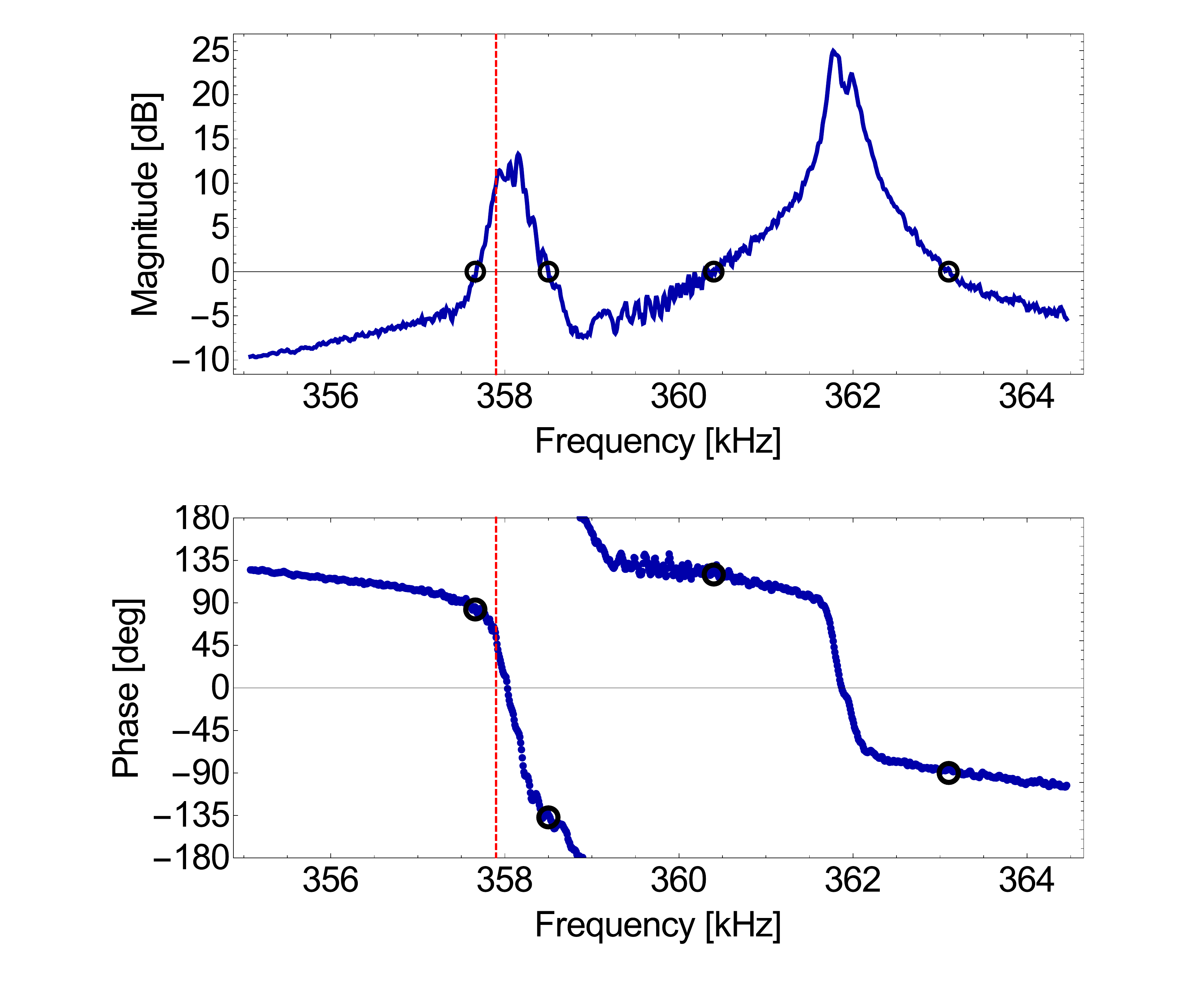}\\
	\caption{Open Loop Transfer function across the two optical modes. Frequency range was scanned using a chirped signal being applied to the EOM. Highlighted in red is the frequency of the mechanical mode. Unity gains are highlighted with black circles, with the worst unity gain phase being 43$\degree$ at 358.5~kHz. In this measurement the corresponding beat note is slightly detuned from the mechanical mode of interest. There is an overall slope in the phase curve of $-12\degree$ per kHz, due to the band-pass filters used in the feedback loop, that have a bandwidth broader than this measurement.}
	\label{fig:tf}
\end{figure}

Fig.~\ref{fig:tf} shows the open loop transfer function for our feedback loop, with the peaks in gain magnitude corresponding to the frequencies of the optical beatings. The open loop gain of the system can be directly measured and represents the product of several aspects of the system, as discussed in Sec.~\ref{sec:analysis}. This measurement was done by adding a swept sine signal from a spectrum analyser to the signal from the QPD, using the spectrum analyser illustrated in Fig.~\ref{fig:expsetup}. By dividing the response of QPD signal by the sum of a swept sine and the QPD signal we make a direct measurement of the open loop gain within a closed loop (as illustrated by \citet{freise03}).

\begin{figure}[h!]
	\centering
	\includegraphics[width=0.65\textwidth]{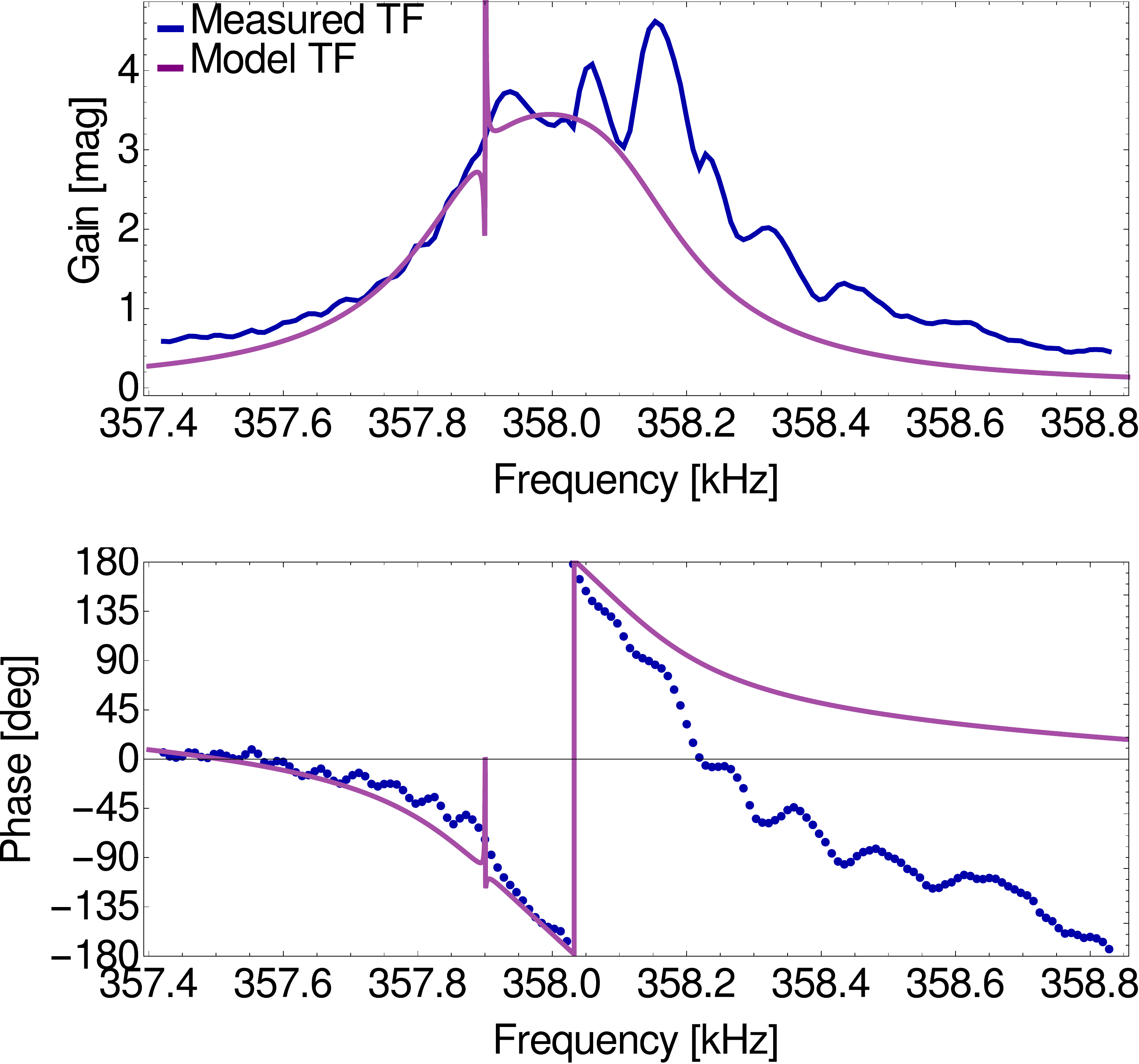}\\
	\caption{Open Loop Transfer Function across the optomechanically unstable mode, illustrating the sharp feature at 357.9~kHz, that could not be seen in the measured data (in blue). The estimated transfer function, in purple, plots plots the closed loop scenario ($a_1' / {a_1}^{\text{probe}}$) frequency shifted to 357.9~kHz, with a detuning of $\Delta_{\text{m}} = - 2 \pi ~ 30$, $G_\text{e} \approx -11$. As the optical beat note would be moving in frequency, due to beam position and thermal state of the cavity changing during a measurement (which consists of the average of many measurements), we observe a disparity between the measurement and this model. Fig.~\ref{fig:tf} shows the full measurement, where the other visible beat note more clearly agrees with theory, though it has no optomechanical interaction.}
	\label{fig:tffit}
\end{figure}

Within the bandwidth of the beating, the phase response is very steep: exhibiting a $180 \degree$ rotation in about 1~kHz. When the frequency of these peaks shifts due to beam position change, thermal effects from optical or $\text{CO}_2$ intensity fluctuations, the frequency of these peaks could change on the order of 1~kHz. This causes our phase margins to change and lead to loop stability being tricky at times. This is highlighted by transfer function measurement issues mentioned in Fig.~\ref{fig:tffit}.

We can attempt to model the observations in Fig.~\ref{fig:tf}, by estimating gain, $G_\text{e}$, and $\Delta_{\text{m}}$ from Eq.~(\ref{eq:openloopmod}). $G_\text{e}$ describes the product of electronic gain, photodiode sensitivity, EOM efficiency, subsequent amplification in the Fibre Laser Amplifier and the unknown coupling coefficient for the fundamental-mode-shaped input sideband to the high order mode within the cavity. $G_\text{e}$ is considered to have a constant gain and phase for all frequencies within the bandwidth of an optical mode. We approximate the losses in the optical modes to be comparable ($\gamma_{1} \approxeq \gamma_0$), and thus we can replicate the observed gain and phase from Fig.~\ref{fig:tf} for the right hand beat note with $G_\text{e} \approxeq -11$ (i.e. with phase of $\pi$), given $R_0 \approxeq 2.5$. Plotting said transfer function (see Fig.~\ref{fig:tffit}) reveals a gain and phase feature for frequencies near the mechanical mode (i.e. less than $\gamma_{ m }$ away from $\omega_{ m }$), when $R_0 > 1$. For large parametric gains, this can manifest as a large dip in the open loop gain, particularly when detuning ($\Delta_{\text{m}}$) is small. A larger loop gain could therefore be required to overcome both instability arising from PI, as well as potentially unstable unity gain frequencies that could arise near the mechanical mode frequency if the PI gain is very large. We do not see this feature in the measured data in Fig.~\ref{fig:tffit} or Fig.~\ref{fig:tf}, since it would lie in a bandwidth of about 2~Hz, and the spacing of the measured data is much too sparse to see it.

\begin{figure}[h!]
	\centering
	\includegraphics[width=0.8\textwidth]{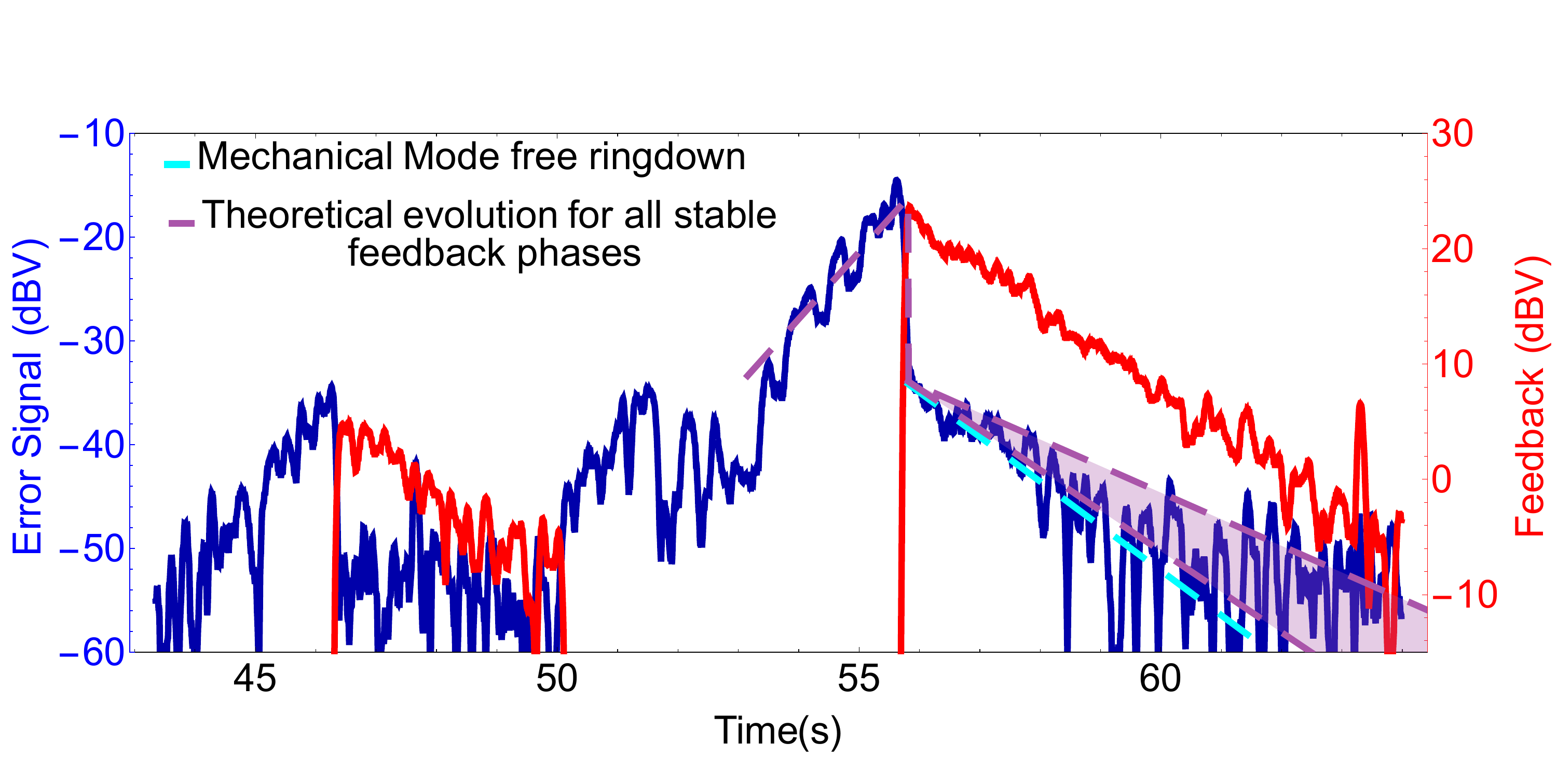}\\
	\caption{Demonstration illustrating damping of observed instability. Here we allowed PI with a Gain of approximately 2.5 to begin ringing up (seen in blue), and enabled the feedback loop (in red) using a switch. The feedback loop was turned on at 56 seconds illustrating a ring-down comparable to that of the free ring-down slope observed in the mechanical mode in the absence of any optical power (cyan line), based on the mechanical Q stated in Sec.~\ref{sec:setupdetails}. The purple lines are the range of theoretical models based on the analysis in Sec.~\ref{sec:analysis}, with time-domain evaluation based on \protect\citet{dan14}. Our recording device appears to have reached a noise floor at about -55~dBV for the error signal, which doesn't affect the feedback loop. The amplitude drop at 56 seconds in the theoretical model is artificial to line up with the sudden suppression of the visibility of the beat notes just as in Fig.~\ref{fig:spect}. The cyan and purple lines are plotted to the same scale as the Error signal.}
	\label{fig:demo}
\end{figure}

\citet{dan14} evaluates a time domain equation for PI evolution. Assuming no knowledge of the feedback phase, we evaluate a reduction of $\text{R}_{0}$ using Eq.~(\ref{eq:pigain_feedback}) for the aforementioned range of stable phases and highlight an area (in purple) in Fig.~\ref{fig:demo} where the best to worst case time evolution scenarios would lie for $|G_\text{e}| = 11$ and an unknown, but stable, feedback phase. The two extremes correspond to an effective reduction of $\text{R}_{0}$ by a factor between 6.2 and 28, for cases where $G_\text{e}$ has a phase of $\pi$ and $\frac{\pi}{2}$ respectively.

Fig.~\ref{fig:demo} demonstrates the effectiveness of the feedback loop to suppress PI as it is being excited with a Parametric Gain of approximately 2.5. Switching on the feedback loop imparts an instantaneous reduction on our signal equivalent to the loop gain at 357.9~kHz, seen at 46 and 56 seconds (this is likely slightly different from what is shown in Fig.~\ref{fig:tf}, since the experiment was conducted on a different day, likely with slightly different cavity power and beam positions). The mechanical mode was excited to a significantly higher amplitude the second time. We see a ringdown of this excitation, which is comparable to the ringdown of the mechanical mode in the absence of optomechanical interaction (free ringdown), though with a longer time constant since the interaction is not completely removed due to finite loop gain and likely sub-optimal feedback phase. As expected, the observation falls somewhere between the two modelled extremes noted in Sec.~\ref{sec:analysis}, as we do not know the exact feedback phase on the day of measurement.

Since we know that we are not injecting a perfectly mode matched field we expect only a small fraction of injected fundamental mode power to couple to a high order mode. This parameter is encompassed in $G_\text{e}$ for Sec~\ref{sec:analysis}, but it is possible to provide an estimate for the percentage of power from sideband which pumps the resonant high order mode. To do so we consider a reduction in high order mode power seen in both peaks in Fig.~\ref{fig:spect} to go from the red to the blue curve (assuming feedback phase is such that it is only destructively interfering, that is: $G_\text{e}$ has a phase of $\pi$), and compare this to the total power which EOM is putting into the sidebands. Comparing these ratios will provide a lower limit estimate for the coupling of our input to the cavity.

To calculate the ratio of sideband power in the cavity injection, we consider the signal applied to the EOM when sustained suppression visible (namely, after 65 seconds in Fig.~\ref{fig:demo}), where a signal with a peak voltage of approximately 400~mV is applied to the EOM. Based on the modulation depth of the EOM that we are using and assuming no further losses, this should produce 45~$\mu$W of power to the modulated sidebands of our 5~W cavity input beam, half of which would couple to the high order mode frequency (only one of the two sidebands), giving an injection power ratio of $4.5 \times 10^{-6}$.

To calculate the change in observable high order mode within the cavity we consider the change in power for the two curves in Fig.~\ref{fig:spect}. Integrating over each curve from 356~kHz to 364.5~kHz we obtain a change in observed power in the beat notes. Given this, its possible to calculate (assuming a constant intra cavity power) the high order mode amplitude ($a_1$) to fundamental mode amplitude ($a_0$) ratio of $a_1/ a_0 = 1.2 \times 10^{-4}$, and thus a power ratio of ${a_1}^2 / {a_0}^2 = 1.4 \times 10^{-8}$.

If we assume that this lower injected sideband power is enough to account for all change in the high order mode, we can estimate a minimum $0.3 \%$ of the injected sideband power couples to our high order modes. This coupling estimates how much of the sideband is essentially pumped into the cavity as a high order mode, and encapsulates the mode-matching\citep{anderson84} of this sideband to the high order mode.

The small coupling ratio, combined with the very small ETM transmission of approximately 25~ppm, means that our feedback loop required very high electronic gain ($4\times 10^{5}$) to achieve the required gain stated in Sec.~\ref{sec:analysis}. Consequently strong filtering was required to prevent small noise sources from saturating our electronics.

For our experiment, we experience no PI when the feedback loop is on constantly, due to the corresponding reduction of effective Parametric Gain (derived in Sec.~\ref{sec:analysis}). This system can be extended to other optical modes which arise from power being scattered out from the fundamental mode, through the addition of more sidebands. Ideally, one needs a sensing array of photo-detectors each sensitive to only a single optical beat mode, and a feedback loop controlling the phase of the corresponding high order mode, so that one does not need to run into the complication of affecting multiple optical modes with a single feedback loop.

\section{Implementation Consideration}\label{sec:imp}

The major advantage of optical feedback compared to mechanical feedback to control parametric instability is reduced complexity. The reduced complexity arises as there are fewer optical modes ($\sim$7 in the arm cavities of aLIGO) compared to unstable mechanical modes ($\sim$20 in each of 4 test masses\citep{gras09} of aLIGO), resulting in a manageable number of potential independent control loops. In this paper effective control has been demonstrated with the simple addition of a frequency modulation feedback.

Implementing such an actuation system in a gravitational wave detector would be a simple task. The power recycling cavity in aLIGO is designed to have a bandwidth of 350~kHz\citep{arain08}. Since all known unstable mechanical modes in aLIGO occur at frequencies well below that, any optical modes that are actuated on would be resonant in the power recycling cavity and would therefore reach the cavity arms where they need to reduce high order mode content. However actuation relies on a fortuitous input coupling between the injected optical mode and the $\text{TEM}_{01}$ mode. Some investigation would be required to determine if this coupling would be adequate in a gravitational wave detector. If this coupling was not large enough some spatial manipulation of the feedback optical field would be required (as previously proposed\citep{fan2010,zhang10}) increasing complexity.

The sensing scheme proposed here has already been demonstrated in mechanical damping experiments at aLIGO\citep{blair17}. In these experiments parametric instability was controlled with mechanical feedback using an error signal generated from the beat note between the fundamental and higher order mode in transmission of the arm cavity as used in this paper. The only foreseeable sensing problem would be poor signal to noise ratio particularly in the case that the optical beat has little overlap with degrees of freedom that can be measured by a QPD. In this case a multi-element photodiode with better overlap with the high order modes may be required.

One potential issue with the proposed scheme arises if suppression is required near the detection band of a gravitational wave detector. aLIGO's 7$^{th}$ order optical modes for example are close to the detection band. In this case feedback filters would have to be more carefully designed to avoid altering the interferometer sensing function or the injection of noise. Fig.~\ref{fig:spect} can illustrate this issue, where some extra noise is injected at frequencies away from the intended actuation bandwidth (contributing extra noise where the blue curve is above the red). Whilst the most sensitive frequency of the detector like aLIGO is around 100~Hz, frequencies as high as 5~kHz are considered important for aLIGO\citep{marty16}. To minimise the effect on interferometer sensitivity one could limit the feedback to a specific frequency around the mechanically excited PI.

The existing control infrastructure already in place for PI feedback through Electro-Static Drives (ESDs)\citep{blair17} at both aLIGO sites and would suit optical feedback design with the addition of broadband frequency modulation.

\section{Conclusion}

We have demonstrated suppression of PI using a novel optical injection scheme capable of alleviating parametric instability (PI) in our cavity. The method involves the feedback of the signal produced from the beating of the fundamental and a high order optical modes sensed at the transmission port of the cavity. An EOM at the injection port generates the necessary frequency matched sidebands, where the lower sideband couples to the high order mode in the cavity to cancel out the high order mode normally produced due to scattering in cavity. 

A major improvement from previous iterations of optical feedback, is that we did not have to produce the correct high order mode shape in order to couple to that mode. Specifically, we estimated the coupling of the generated sideband to the high order mode, to be at least $0.3 \%$, which is still much larger than power ratio of the high order mode to the fundamental mode within the cavity. This makes utilising this feedback system easier and more robust against alignment issues than previous proposals.

In Fig.~\ref{fig:demo}, we demonstrate a dramatic suppression of the Parametric Gain on an unstable mode, with a gain reduction of at between 6.2 and 28 times, as derived in Sec.~\ref{sec:analysis}. This demonstrates this scheme's ability to suppress high gain PI, which can be further improved with higher electronic feedback gain. Furthermore we demonstrated that continual application of this feedback would prevent the onset of instability.

In our experiment we demonstrated the suppression of two high order modes, and it is possible to sum signals produced by any number of high order mode beat notes. This would allow suppression of PI produced by any number of optical modes, remedying PI for all mechanical modes affected by that set of beat notes. aLIGO already has infrastructure capable of sensing the required signals and implementing this scheme is already possible.

\section*{Acknowledgements}
We wish to thank the LIGO Scientific Collaboration Advanced Interferometer Configurations and Optics Working Group for useful advice. This work is supported by Australian Research Council Discovery Project DP160102447, and Center of Excellence for Gravitational Wave Discovery project CE170100004. 

The authors would also like tho thank Joris van Heijningen and Jue Zhang for useful input into countless aspects of this paper. We are also thankful Steve Key and John Moore who contribute so much to the maintenance and technical improvements made to the Gingin site.



\begin{thebibliography}{23}
\providecommand{\natexlab}[1]{#1}
\providecommand{\url}[1]{\texttt{#1}}
\expandafter\ifx\csname urlstyle\endcsname\relax
  \providecommand{\doi}[1]{doi: #1}\else
  \providecommand{\doi}{doi: \begingroup \urlstyle{rm}\Url}\fi

\bibitem[{S}cientific {C}ollaboration(2015)]{ligo15}
The~{LIGO} {S}cientific {C}ollaboration.
\newblock {A}dvanced {LIGO}.
\newblock \emph{Classical and Quantum {G}ravity}, 32\penalty0 (7):\penalty0
  074001, 2015.

\bibitem[Acernese et~al.(2015)Acernese, Agathos, Agatsuma, Aisa, Allemandou,
  Allocca, Amarni, Astone, Balestri, Ballardin, et~al.]{acernese2015}
F~Acernese, M~Agathos, K~Agatsuma, D~Aisa, N~Allemandou, A~Allocca, J~Amarni,
  P~Astone, G~Balestri, G~Ballardin, et~al.
\newblock {A}dvanced {V}irgo: a {S}econd-{G}eneration {I}nterferometric
  {G}ravitational {W}ave {D}etector.
\newblock \emph{Classical and Quantum {G}ravity}, 32\penalty0 (2):\penalty0
  024001, 2015.

\bibitem[Aso et~al.(2013)Aso, Michimura, Somiya, Ando, Miyakawa, Sekiguchi,
  Tatsumi, and Yamamoto]{aso13}
Yoichi Aso, Yuta Michimura, Kentaro Somiya, Masaki Ando, Osamu Miyakawa,
  Takanori Sekiguchi, Daisuke Tatsumi, and Hiroaki Yamamoto.
\newblock Interferometer design of the kagra gravitational wave detector.
\newblock \emph{Phys. Rev. D}, 88:\penalty0 043007, Aug 2013.
\newblock \doi{10.1103/PhysRevD.88.043007}.
\newblock URL \url{https://link.aps.org/doi/10.1103/PhysRevD.88.043007}.

\bibitem[Braginsky et~al.(2001)Braginsky, Strigin, and Vyatchanin]{brag01}
V.~Braginsky, S.~Strigin, and S.~P. Vyatchanin.
\newblock {P}arametric {O}scillatory {I}nstability in {F}abry-{P}erot
  {I}nterferometer.
\newblock \emph{Physics Letters A}, 287\penalty0 (5):\penalty0 331--338, 2001.

\bibitem[Zhao et~al.(2015)Zhao, Ju, Fang, Blair, Qin, Blair, Degallaix, and
  Yamamoto]{zhao2015}
Chunnong Zhao, Li~Ju, Qi~Fang, Carl Blair, Jiayi Qin, David Blair, Jerome
  Degallaix, and Hiroaki Yamamoto.
\newblock {P}arametric {I}nstability in long optical cavities and suppression
  by dynamic transverse mode frequency modulation.
\newblock \emph{Physical Review D}, 91\penalty0 (9):\penalty0 092001, 2015.

\bibitem[Evans et~al.(2015)Evans, Gras, Fritschel, Miller, Barsotti, Martynov,
  Brooks, Coyne, Abbott, Adhikari, et~al.]{evans2015}
Matthew Evans, Slawek Gras, Peter Fritschel, John Miller, Lisa Barsotti, Denis
  Martynov, Aidan Brooks, Dennis Coyne, Rich Abbott, Rana~X Adhikari, et~al.
\newblock Observation of {P}arametric {I}nstability in advanced {LIGO}.
\newblock \emph{Physical Review Letters}, 114\penalty0 (16):\penalty0 161102,
  2015.

\bibitem[Blair et~al.(2015)Blair, Ju, Zhao, Wen, Miao, Cai, Gao, Lin, Liu, Wu,
  et~al.]{blair2015}
David Blair, Li~Ju, Chunnong Zhao, LinQing Wen, HaiXing Miao, RongGen Cai,
  JiangRui Gao, XueChun Lin, Dong Liu, Ling-An Wu, et~al.
\newblock {T}he next detectors for {G}ravitational {W}ave astronomy.
\newblock \emph{Science China Physics, Mechanics \& Astronomy}, 58\penalty0
  (12):\penalty0 1--34, 2015.

\bibitem[Degallaix et~al.(2007)Degallaix, Zhao, Ju, and Blair]{degallaix2007}
J{\'e}r{\^o}me Degallaix, Chunnong Zhao, Li~Ju, and David Blair.
\newblock {T}hermal tuning of optical cavities for {P}arametric {I}nstability
  control.
\newblock \emph{JOSA B}, 24\penalty0 (6):\penalty0 1336--1343, 2007.

\bibitem[Blair et~al.(2017)Blair, Gras, et~al.]{blair17}
Carl Blair, Slawek Gras, et~al.
\newblock First demonstration of electrostatic damping of parametric
  instability at advanced ligo.
\newblock \emph{Phys. Rev. Lett.}, 118:\penalty0 151102, Apr 2017.
\newblock \doi{10.1103/PhysRevLett.118.151102}.
\newblock URL \url{https://link.aps.org/doi/10.1103/PhysRevLett.118.151102}.

\bibitem[Gras et~al.(2015)Gras, Fritsche, Barsotti, and Evans]{gras15}
S.~Gras, P.~Fritsche, L.~Barsotti, and M.~Evans.
\newblock Resonant dampers for parametric instabilities in gravitational wave
  detectors.
\newblock \emph{Physical Review D}, 92\penalty0 (8):\penalty0 082001, 2015.

\bibitem[Gras et~al.(2009)Gras, Blair, and Zhao]{gras2009}
S~Gras, DG~Blair, and C~Zhao.
\newblock {S}uppression of parametric instabilities in future {G}ravitational
  {W}ave detectors using damping rings.
\newblock \emph{Classical and Quantum {G}ravity}, 26\penalty0 (13):\penalty0
  135012, 2009.

\bibitem[Fan et~al.(2010)Fan, Merrill, Zhao, Ju, Blair, Slagmolen, Hosken,
  Brooks, Veitch, and Munch]{fan2010}
Yaohui Fan, Lucienne Merrill, Chunnong Zhao, Li~Ju, David Blair, Bram
  Slagmolen, David Hosken, Aidan Brooks, Peter Veitch, and Jesper Munch.
\newblock {T}esting the suppression of opto-acoustic parametric interactions
  using optical feedback control.
\newblock \emph{Classical and Quantum {G}ravity}, 27\penalty0 (8):\penalty0
  084028, 2010.

\bibitem[Miller et~al.(2011)Miller, Evans, Barsotti, Fritschel, MacInnis,
  Mittleman, Shapiro, Soto, and Torrie]{miller2011}
John Miller, Matthew Evans, Lisa Barsotti, Peter Fritschel, Myron MacInnis,
  Richard Mittleman, Brett Shapiro, Jonathan Soto, and Calum Torrie.
\newblock {D}amping parametric instabilities in future {G}ravitational {W}ave
  detectors by means of electrostatic actuators.
\newblock \emph{Physics Letters A}, 375\penalty0 (3):\penalty0 788--794, 2011.

\bibitem[Zhang et~al.(2010)Zhang, Zhao, Ju, and Blair]{zhang10}
Zhongyang Zhang, Chunnong Zhao, Li~Ju, and David Blair.
\newblock Enhancement and suppression of opto-acoustic parametric interactions
  using optical feedback.
\newblock \emph{Physical Review A}, 81\penalty0 (1):\penalty0 013822, 2010.

\bibitem[Danilishin et~al.(2014)Danilishin, Vyatchanin, Blair, Li, and
  Zhao]{dan14}
Stefan~L. Danilishin, Sergey~P. Vyatchanin, David~G. Blair, Ju~Li, and Chunnong
  Zhao.
\newblock Time evolution of parametric instability in large-scale
  gravitational-wave interferometers.
\newblock \emph{Physical Review D}, 90\penalty0 (11):\penalty0 122008, 2014.

\bibitem[Zhao et~al.(2008)Zhao, Ju, Fan, Gras, Slagmolen, Miao, Barriga, Blair,
  Hosken, Brooks, et~al.]{zhao2008}
C~Zhao, Li~Ju, Y~Fan, S~Gras, Bram~JJ Slagmolen, Hong Miao, P~Barriga, David~G
  Blair, David~John Hosken, Aidan~Francis Brooks, et~al.
\newblock {O}bservation of three-mode parametric interactions in long optical
  cavities.
\newblock \emph{Physical Review A}, 78\penalty0 (2):\penalty0 023807, 2008.

\bibitem[Drever et~al.(1983)Drever, Hall, Kowalski, Hough, Ford, Munley, and
  Ward]{drever83}
R.W.P. Drever, John Hall, F~Kowalski, James Hough, G.M. Ford, A.J. Munley, and
  Hywel Ward.
\newblock Laser phase and frequency stabilization using an optical resonator.
\newblock \emph{Applied Physics B}, 31:\penalty0 97--105, 06 1983.

\bibitem[Fang et~al.(2021)Fang, Blair, Zhao, and Blair]{fang21}
Qi~Fang, Carl~D. Blair, Chunnong Zhao, and David~G. Blair.
\newblock Revealing optical loss from modal frequency degeneracy in a long
  optical cavity.
\newblock \emph{Opt. Express}, 29\penalty0 (15):\penalty0 23902--23915, 2021.

\bibitem[Anderson(1984)]{anderson84}
Dana~Z. Anderson.
\newblock Alignment of resonant optical cavities.
\newblock \emph{Applied Optics}, 23\penalty0 (17):\penalty0 2944--2949, 1984.

\bibitem[Freise(2003)]{freise03}
Andreas Freise.
\newblock \emph{The {N}ext {G}eneration of {I}nterferometry:
  {M}ulti-{F}requency {O}ptical {M}odelling, {C}ontrol {C}oncepts and
  {I}mplementation}.
\newblock PhD thesis, Hannover : Universität, 2003.

\bibitem[Gras(2009)]{gras09}
Slawomir~M. Gras.
\newblock \emph{{O}pto-acoustic {I}nteractions in {H}igh {P}ower
  {I}nterferometric {G}ravitational {W}ave {D}etectors}.
\newblock PhD thesis, The University of Western Australia, 2009.

\bibitem[Arain and Mueller(2008)]{arain08}
Muzammil~A. Arain and Guido Mueller.
\newblock Design of the {A}dvanced {LIGO} recycling cavities.
\newblock \emph{Optics Express}, 16\penalty0 (14):\penalty0 10018, 2008.

\bibitem[Martynov et~al.(2016)Martynov, Hall, et~al.]{marty16}
D.~V. Martynov, E.~D. Hall, et~al.
\newblock Sensitivity of the {A}dvanced {LIGO} detectors at the beginning of
  gravitational wave astronomy.
\newblock \emph{Phys. Rev. D}, 93\penalty0 (11):\penalty0 112004, Jun 2016.

\end{thebibliography}

\end{document}